# Instrumentation for the detection and characterization of exoplanets

Francesco Pepe[1], David Ehrenreich[1], Michael R. Meyer[2]

[1]Observatoire Astronomique de l'Université de Genève, 51 ch. des Maillettes, 1290 Versoix, Switzerland
[2]Swiss Federal Institute of Technology, Institute for Astronomy, Wolfgang-Pauli-Strasse 27, 8093 Zurich, Switzerland

**In no other field of astrophysics has the impact of new instrumentation been as substantial as in the domain of exoplanets. Before 1995 our knowledge about exoplanets was mainly based on philosophical and theoretical considerations. The following years have been marked, instead, by surprising discoveries made possible by high-precision instruments. More recently the availability of new techniques moved the focus from detection to the characterization of exoplanets. Next-generation facilities will produce even more complementary data that will lead to a comprehensive view of exoplanet characteristics and, by comparison with theoretical models, to a better understanding of planet formation.**

Astrometry is the most ancient technique of astronomy. It is therefore not surprising that the first (unconfirmed) detection of an extra-solar planet arose through this technique[1]. In 1984, another detection of a planetary-mass object around the nearby star VB 8 was claimed, this time using speckle interferometry[2], but subsequent attempts to locate it were unsuccessful. It was finally Doppler velocimetry that delivered the first unambiguous detection of a probable brown dwarf around HD 114762[3]. In 1992, a handful of bodies of terrestrial mass were found[4] and confirmed, by the measurement of timing variation, to orbit the pulsar PSR1257+12. Although very powerful, this technique was restricted to a small number of very particular hosts. Doppler velocimetry, instead, could be applied with good results to basically any 'quiet' star showing a reasonable amount of narrow absorption lines in its spectrum. The continuous improvement of this technique led, in 1995, to the discovery of the first giant planet around the Sun-like star 51 Pegasi[5] and marked the start of an intensive era of discoveries (see also the paper by Mayor et al. in this same journal issue).

Since the discovery of 51 Peg b, microlensing, transit searches and direct imaging delivered, together with Doppler velocimetry, an increasing number of planets and planetary candidates. Better instruments and improved detection limits push towards the detection of low-mass and small planets. The discovery of multi-planetary systems is furthermore the direct consequence of long-term, high-precision programs. A new breakthrough was made thanks to space-based transits searches *CoRoT*[6] and *Kepler*[7]. These missions significantly contributed to the statistical study of the exoplanetary systems.

In this Review, we will discuss techniques and instruments that have most contributed to our understanding of exoplanets. We will also provide an overview of present and future instrumentation and describe how the field is moving from simple detection and statistical studies to the characterization of individual planets, their interior and their atmospheric composition.

## Stellar radial velocities

Giant planets on short orbits induce radial-velocity variations of their host stars of several tens to a few hundreds of meters per second. Early Doppler velocimeters[8,9] delivered 200–500 m s$^{-1}$ precision. With the use of a hydrogen fluoride (HF) absorption cell the precision could be improved by one order of magnitude[10]. In the late eighties and early nineties an entire suite of new techniques and spectrographs[11-14] led to an improvement of the radial-velocity precision down to 3 to 15 m s$^{-1}$. The better precision led in turn to the discovery of 51 Peg b[5] and the era of giant-planet detection.

Would it be possible to detect terrestrial-mass exoplanets by the Doppler technique? Some astronomers believed that improving the instrumental precision would be a key element[15]. Confirmation of this view was provided by the discovery of μ Ara c in 2004[16]: Of only 10 times the mass of the Earth and an orbit of 9.6-days, this planet produces a stellar radial-velocity pull of 3 m $s^{-1}$ semi-amplitude. The detection of this tiny signal required a new generation of spectrographs such as *HARPS*[18]. It represented the first step towards the detection and characterization of a vast population of Neptune-mass planets and super-Earths. The longer the temporal coverage and the better the instrumental precision, the smaller radial-velocity signals could be detected (Figure 1).

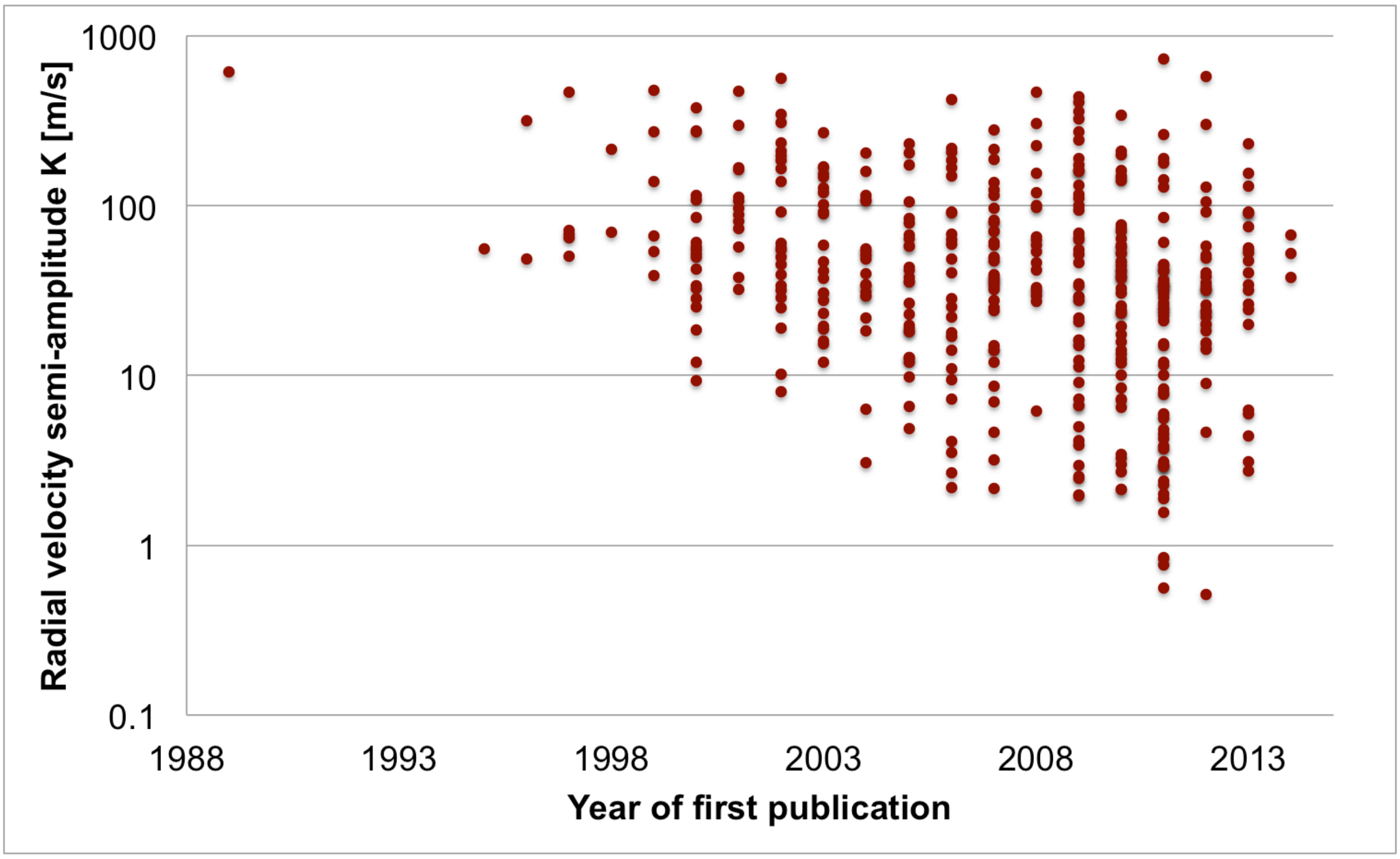

**Figure 1:** ***Radial-velocity semi-amplitude of planetary-mass companions.*** **All planets discovered by the Doppler technique from 1989 to present days are plotted as a red dot. Remarkably, the detection limit improved by three orders of magnitudes in less than 3 decades. Underlying data were retrieved from www.exoplanets.org.**

The Doppler measurement consists in determining the wavelength of an identified spectral line and comparing it with the theoretical value it would have when transferred into the solar system's rest frame. The Doppler equation links the measurement to the theoretical wavelength via the relative velocity vector, finally delivering the projection of this vector in the direction of the line of sight (radial velocity). In order to increase the precision, the average over several thousands of spectral lines is computed. It should be noted, however, that the radial-velocity measurement is affected by several potential error sources that have been discussed extensively[10,14,20,21]. The main error sources are: photon noise[14,22], instrumental errors[11,14,20], spectrograph illumination effects[23,24], spectral contamination[20,25], and stellar 'noise'[26-40], commonly referred to as stellar jitter. The term stellar jitter masks various stellar causes that produce radial-velocity effects at all time scales and of different magnitude. The discussion of all these effects lies again beyond the purpose of our review. Nevertheless, it shall be reminded that stellar jitter is probably the strongest limitations for Doppler velocimetry when aiming at sub-meter-per-second precision.

Present and future Doppler spectrographs will have to address the mentioned limitations. As a first step it will be necessary to increase the telescope size, since high spectral resolution measurements are photon-starved, even for relatively bright targets. The gain obtained with a large telescope is however easily lost if spectral resolution is low. In fact, for unresolved spectral lines, the measurement precision increases significantly with increasing the spectral resolution[22]. In the photon-noise limited regime the error $\varepsilon_{\mathrm{bary}}$ on the line-centre measurement can be estimated to

$$\varepsilon_{\mathrm{bary}} = \frac{\sqrt{\sigma^2}^{1.5}}{\sqrt{2 \cdot I_0 \cdot EW}} \cdot \sqrt{\left(1 - \frac{c}{2}\right)},$$

where $\sigma$ is the measured width of the spectral line as seen through the spectrograph, $c = (I_{min} - I_0)/I_0$ is the measured line contrast and $EW = \sigma\, c$ the equivalent width. $I_0$ and $I_{\mathrm{min}}$ designate the photoelectron counts per resolution element in the continuum and the line minimum, respectively. It must be noted that the resolution element can be represented either by the detector pixel or the wavelength unit as long as all the parameters are expressed in the same units. It is commonly agreed nowadays that a spectral resolution of at least $R := \lambda/\Delta\lambda = 100{,}000$ should be used in order to guarantee the best precision on slowly-rotating, quiet, solar-type stars. Spectral resolution and adequate line sampling not only allow us to achieve better signal-to-noise per spectral line, but also to reduce possible instrumental errors in both the radial-velocity measurement and the calibration process. To first order, instrumental errors scale with the size of the resolution element (expressed in wavelength units). Unfortunately, with increasing telescope size, spectral resolution is a considerable cost driver. For seeing-limited instruments the optical etendue $E$ ($E = A \times \Omega$, i.e. the beam cross-section area times the solid angle) increases with the telescope size, and so does the instrument size if the spectral resolution is kept fixed[17]. In the era of 8-m class and extremely-large telescopes (ELTs), this aspect has become a technical and managerial challenge that is nevertheless successfully addressed by employing novel optical design concepts[193-193].

All the future projects for radial-velocity spectrographs (Table 1) aim at detecting rocky planets in the habitable zone (HZ)[42] of a solar-like and low-mass star (a distance to the star at which liquid water can persist on the surface of the planet). In order to attain this objective they must be photon-efficient and precise to the sub meter-per-second level. Photon efficiency is obtained with optimized designs and high spectral resolution. High precision requires also the control of all instrumental effects. It has therefore become state-of-the-art to design stable instruments[41]. Gravity invariance and illumination stability of the spectrograph are critical aspects that can only be obtained through a fibre feed[43,44]. Despite the intrinsic light scrambling properties of optical fibres[45-47] it was soon realised that the illumination produced by a circular optical fibre depends on how the star is fed into the fibre. In other words, motions of the stellar image at the fibre entrance would produce a change in the illumination of the spectrograph and mimic a radial velocity effect. A considerable effort was invested in improving image scrambling by employing double scramblers[45,13] and octagonal fibres[48,49]. Effective improvements have already been demonstrated on operational instruments[50,51].

Any kind of effect that introduces a distortion or a shift of the spectral line in the detector-pixel space will be interpreted, if not perfectly monitored, as a wavelength change and thus a Doppler shift[20]. Two methods of tracking the instrumental-profile changes have successfully been applied in the past: The first is to superimpose an absorption spectrum of a reference gas cell[10,14,52] on the stellar spectrum, such that the instrumental profile (IP) is continuously measured. The so-called self-calibration technique is particularly useful and effective in spectrographs with varying instrument profile, as in the case of slit spectrographs. The disadvantage of this technique is the restricted bandwidth of the gas-cell spectrum, the loss of efficiency due to absorption in the light path, and the necessity for a sophisticated de-convolution process in order to recover the stellar spectrum and thus the radial velocity. This latter step requires the introduction of many additional parameters for spectral modelling. In order to obtain a given precision higher signal-to-noise spectra must be acquired. The second method, the so-called simultaneous reference technique[13,18], is conceptually opposite. It assumes a stabilized IP that does not change between two wavelength-calibrations of the spectrograph, such that the determined relation between the detector pixel and the wavelength remains valid over these time scales (typically a night). A second channel carrying a spectral reference is continuously fed to the spectrograph to monitor and correct for potential instrumental drifts or IP changes. It must be guaranteed, however, that the changes suffered by the scientific and the reference channels are identical over timescales of one observing night. Therefore, the whole design of the instrument must be optimized for stability, requesting fibre feed and light scrambling, as well as pressure, mechanical, thermal and optical stability. The effort is compensated by an unrestricted spectral bandwidth and the acquisition of an 'uncontaminated' scientific spectrum.

Although, in the case of the self-calibration technique, the instrument profile is supposed to be recoverable by de-convolution, there seems to be general agreement on the fact that low-order instrument-profile changes must be avoided in any case and that a stable instrument will eventually deliver more precise measurements. There is also agreement on the need for better calibration sources. The laser-frequency comb[53-58], when available at full potential, will provide the aimed calibration accuracy and precision. In the meantime, alternative sources are being developed, as for example passive Fabry-Pérot cavities[59-61] for simultaneous reference, or actively stabilized Fabry-Pérot systems for wavelength calibration[62].

The near-infrared (NIR) wavelength region is becoming increasingly interesting because of two other aspects: First, M-dwarfs are much brighter in the infrared than in the visible[63]. These stars are cooler and thus their Habitable Zone lies closer to the host star. In addition the parent star is less massive. Potential habitable planets are therefore more easily detected by radial-velocity[63]. The second advantage of the NIR compared to the visible is that the influence of spots is strongly reduced[64-66]. A comparison with the radial-velocity determined in the visible wavelength range may furthermore help in discriminating a planet-induced velocity change from a stellar effect. For all these reasons many new instruments[67-72] (see also Table 1) will operate in the infrared wavelength domain. The use of adaptive optics[73] might be a mean to reduce size and costs of these instruments.

**Transit photometry and spectroscopy**

There are two approaches to detect planetary transits: (i) surveying as many stars as possible with one or several photometers in hope to detect new exoplanets through their transits, and (ii) photometrically following-up planets discovered by Doppler velocimetry around their predicted inferior conjunction time. (The inferior conjunction denotes the orbital configuration where the planet lies between its host star and the observer; a transit occurs at the inferior conjunction if the orbital plane of the planet is aligned with the line of sight.) In the first method, the expected depth of the transit light curve dictates the photometric precision needed: for Jupiter-size planets in transit across Sun-like stars, the transits can be detected from the ground with amateur telescopes. Hot jupiters, however, are only found orbiting about 1% of nearby solar-type stars[74], requiring observers to maximise the number of surveyed stars. Bright main sequence stars can be surveyed over large fraction of the sky by wide-field cameras with small aperture telescopes and charge-coupled devices (CCDs), as illustrated by the *Wide Angular Search for Planets* (*WASP*[75]). Observations from a single location are limited however by the duration of the night. Time and sky coverage can be further improved with networks of small telescopes relaying from different longitudes, such as the *Hungarian Automated Telescope Network* (*HATNet*[76]) or the *Trans-Atlantic Exoplanet Survey* (*TrES*[77]).

The other strategy is to stare at crowded stellar fields. The 1.3-m telescope of *the Optical Gravitational Lensing Experiment* (*OGLE*) yielded the first discoveries of exoplanets through the transit method[78] by applying this strategy. The confirmation of these detections with velocimetry[79] however required a large observational effort because of the faint optical magnitudes (denoted $V$) of the stars surveyed ($14 < V < 16$). The first space missions dedicated to search for transiting exoplanets, *CoRoT (Convection, Rotation, and planetary Transits*[6,80]*)* and *Kepler*[7], also stared at dense fields with high-cadence precise (relative) photometry (see the reviews by Hatzes and Lissauer et al. in this journal issue). Together, these satellites have surveyed several hundred thousands of stars. Radial-velocity follow-up of *CoRoT* and *Kepler* exoplanet candidates remained difficult due to the faint magnitudes of the host stars and the large number of targets needing follow-up. The faintness of the host stars set severe limits also to the use of photon-starved techniques, such as transmission spectroscopy for the study of the planetary atmospheres. This technique requires bright host stars (see Figure 2), such as the hosts of planets discovered through velocimetry and later detected in transit. Only nine such exoplanets are known so far, but future space missions will search for more planets transiting bright stars. In the meantime, and from ground, it is being looked for planets transiting small stars such as M dwarfs, since the transit signal is inversely proportional to the square of the stellar radius. The *Mearth* survey[81], composed by eight identical robotically-controlled 40-cm telescopes with CCD detectors, found a super-Earth[82] especially amenable to follow-up atmospheric studies,[83-87].

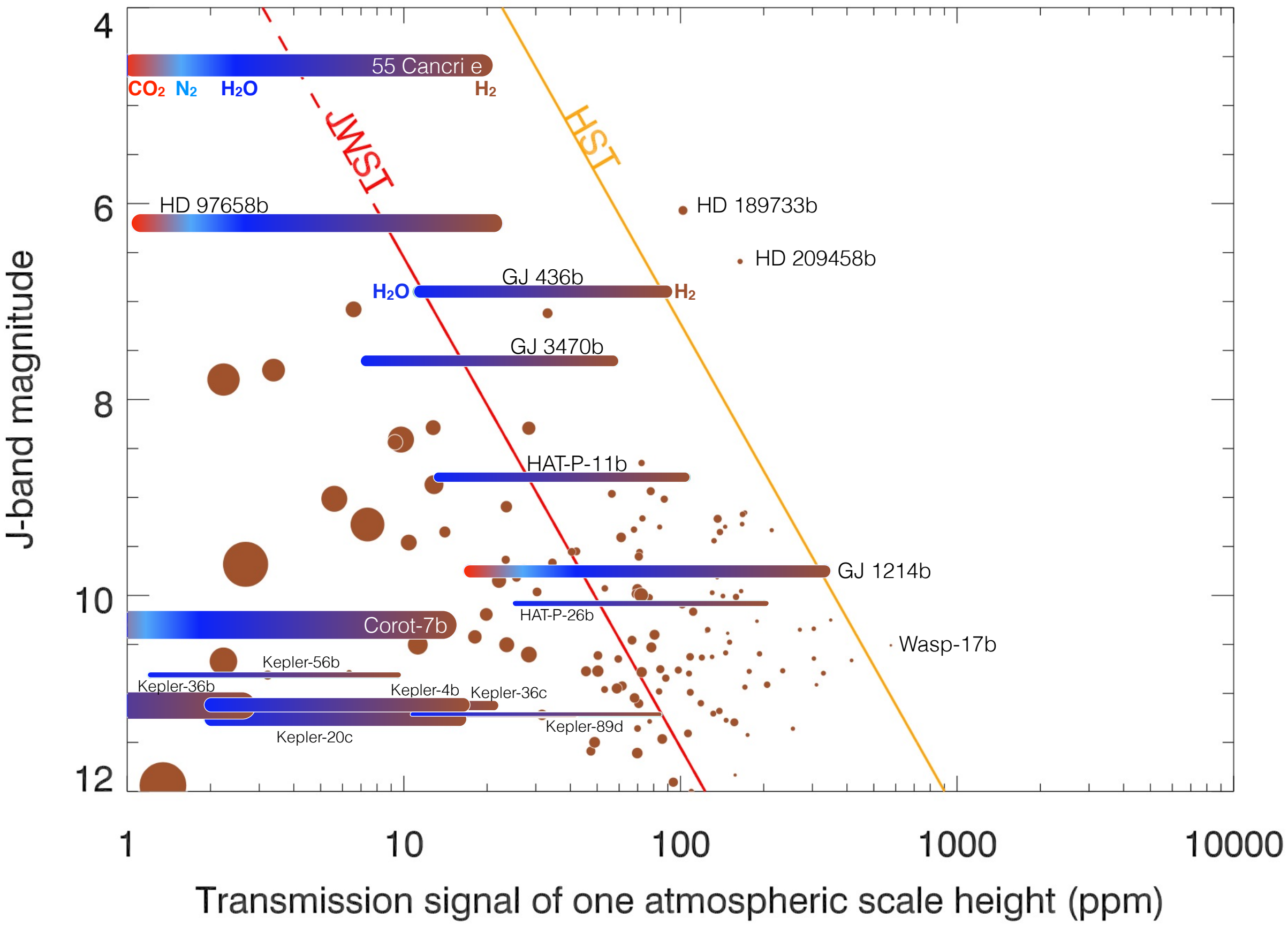

**Figure 2:** ***Detectability of planetary atmospheres.*** **The signal of one atmospheric scale height seen in transmission during transit is plotted versus the stellar *J* magnitude. The signal is calculated in p.p.m. as $2{\times}10^6$ ($\Delta F/F$) ($H/R_p$), where $\Delta F/F$ is the transit depth and $H$ is the atmospheric scale height. This quantity scales here with the equilibrium temperature of the planet and is inversely proportional to the acceleration of gravity at the surface of the planet and the mean molar mass ($\mu$) of the atmosphere. The atmospheric signal is proportional to the planet mean density. The size of the circles (and the thickness of the colour bars) scales with the density to show this effect. For giant exoplanets (brown circles), the atmosphere is assumed primarily composed of molecular hydrogen ($H_2$) and helium ($\mu = 2.3$ g mol$^{-1}$). For lower-mass planets, such as neptunes ($10 < M_p < 60$ M$_\oplus$) and super-earths ($M_p < 10$ M$_\oplus$), the atmospheric composition is unknown and the colour bar extents represent all possible signal values assuming hydrogen/helium- ($\mu = 2.3$ g mol$^{-1}$, brown) and water- ($H_2O$, $\mu = 18$ g mol$^{-1}$, blue) dominated atmospheres for neptunes, and molecular nitrogen- ($N_2$, $\mu = 28$ g mol$^{-1}$, sky blue), and carbon dioxide- ($CO_2$, $\mu = 44$ g mol$^{-1}$, red) dominated atmospheres for super-earths, in addition to the two earlier types. Approximate *HST* and *JWST* 3-σ detection limits (orange and red lines, respectively) are shown. Only super-earths and neptunes with a mass determined to better than 20% are represented.**

## Studies of exoplanetary atmospheres

The hot gas giant HD 209458b was the first exoplanet captured in transit separately by two small telescopes[88,89] with relative photometric precision of 0.2–0.4%. This transit was also the first exoplanet-related event observed from space: the 2.4-m *Hubble Space Telescope (HST)* measured the transit light curve to a precision of 110 ppm per minute of observation[90]. The photometric observations of HD 209458b were obtained by integrating the stellar spectra collected before, during and after the transit by the *Space Telescope Imaging Spectrograph* (*STIS*[91]) CCD detector. These spectra were recorded with a medium-resolution ($R := \lambda/\Delta\lambda = 5540$) grism of medium band pass, notably including the sodium doublet at 589 nm. The first transmission signature of an exoplanetary atmosphere was reconstructed from this data set by measuring, during the transit, an extra absorption of 200 ppm in the sodium lines[92]. The far-ultraviolet channel of the *STIS* instrument, which collects ultra-violet photons with a multi-anode microchannel array (MAMA) detector, was employed to observe the transit of HD 209458b over the stellar Lyman-α emission of atomic hydrogen at 121 nm. These measurements led to the discovery of an extended upper atmosphere to the planet[93].

HD 209458b remained for quite some time the only known transiting exoplanet. By the time additional transiting exoplanets were announced (2004), *STIS* had experienced a power supply failure. The instrument was only repaired in 2009 during the last servicing mission of *HST*. Arguably, the main effect of the *STIS* failure was to shift the field of exoplanetary atmospheres into the infrared. After 2004, and in spite of successful attempts to record precise transit light curves with the *Advanced Camera for Surveys* on-board of *HST*[94], the 85-cm *Spitzer Space Telescope* became the prime observatory not only for transits, but also for eclipses of planets by their stars, which can occur at superior conjunctions (the orbital configuration opposite to the inferior conjunction, when the planet passes behind the star). Broad-band photometry of these eclipses with the *Infrared Camera for Survey* (*IRAC*[95]) on *Spitzer* revealed the thermal emission from exoplanets, the first example of direct detection of light from a planet orbiting a star[96-98]. The instrument has four broad-band infrared channels collecting light on two detectors made of indium-antimonide (InSb, 3.6 and 4.5 μm channels) and arsenic-doped silicon (Si:As, 5.8 and 8.0 μm channels).

The first infrared observation of a planetary transit[99] was obtained with the *Multiband Imaging Photometer for Spitzer* (*MIPS*) at 24 μm. These observations were limited by the low stellar flux in the mid-infrared. Furthermore, transit observations in the near infrared exhibited large instrumental effects, precluding the detection of molecular signatures. Both photometry with *IRAC*[100-102] and spectroscopy with the *Near-Infrared Camera and Multi-Object Spectrometer* (*NICMOS*[103]) on *HST* yielded non-reproducible results or were of insufficient quality for unambiguous interpretation[104-106]. Eclipse spectroscopy of the dayside emission of HD 189733b obtained with the third instrument on *Spitzer*, the *Infrared Spectrograph* (*IRS*[107]) providing low-resolution ($R$ = 80) and spectral coverage from 5 to 14 μm, also had to be corrected for instrumental effects[108]. The *IRS* data nonetheless provided evidence for molecular absorption in an exoplanet atmosphere[109]. Unfortunately, the use of *IRS* was terminated after *Spitzer* ran out of cryogen on May 2009. Meanwhile, *Spitzer* continues observing with *IRAC* 3.6- and 4.5-μm channels, now commonly used to obtain precise transit light curves of exoplanets down to the super-Earth size regime[110,111].

Ground-based atmospheric characterisation of exoplanets advanced through the use of high-resolution spectrographs. The signature of sodium in the atmosphere of HD 209458b was retrieved[112] in data taken with the *High Dispersion Spectrograph* ($R$ = 45,000) at the Subaru 8-m telescope[113]. The employed technique, differential spectroscopy, consists in calibrating the signal in the spectroscopic features by the continuum signal in the vicinity of the features. The "absolute" transit depth is lost, but the transmission signal can be retrieved assuming that telluric absorption can be sufficiently calibrated. Another technique is to calibrate the wavelength-dependent signal using other stars within the field of view of the instrument. This can be achieved in spectrophotometry for systems with nearby reference stars[114,115] or in spectroscopy with slit masks positioned on the target and on several reference stars in the field{Bean:2010ct}. A break-through was made possible by the *Cryogenic High-Resolution Infrared Echelle Spectrograph* (*CRIRES*) on the Very Large Telescope (VLT): its high resolution ($R$ = 100,000), although over a narrow (50 nm) wavelength infrared region, allows tracking the wavelength shift of individual spectral features composing molecular bands of water, carbon monoxide, or carbon dioxide present in the atmosphere of the planet as the planet orbits the star[116-118]. The method works for transiting and non-transiting planets alike, giving access to the brightest exoplanetary systems, such a τ Boötis[119]. Its application to the directly imaged planet β Pictoris b[120] led to the determination of the spin velocity of the planet[121].

The refurbishment of *HST* in May 2009 enabled the recovery of *STIS* capabilities and the start of operations of both the *Cosmic Origins Spectrograph* (*COS*[122]) in the far-ultraviolet and the *Wide-Field Camera-3* (*WFC3*[123]) in the near-infrared. *COS* and *STIS* provided observations and confirmation of the atmospheric mass loss from HD 209458b in the singly ionised carbon lines at 133 nm[124]. These measurements were extended to other exoplanets[125-129]. Visible STIS spectra revealed atomic signatures and the presence of light scattering processes in the upper atmospheric layers of HD 189733b[125, 130], and observations of the eclipse of HD 189733b with a low-resolution grating from 290 to 570 nm also yielded the first chromatic measurements of a planetary albedo[131]. The WFC3 infrared channel was successfully used for slitless grism spectroscopy of exoplanetary transits in the near-infrared, achieving near-photon-noise transmission spectroscopy of super-Earths, Neptunes, and gas giants between 1.1 and 1.7 µm[86,132,-139], and detecting the 1.38-µm water band in some of these planets.

**Direct imaging and astrometry**

Despite many years of technological development, the search for ideal targets, improved analysis algorithms, and investment of observing time on leading telescopes, it was not until 2008 that the first direct images of an extra-solar planetary system around a star were obtained. The multi-planet system HR 8799 with all planets[140] orbiting the intermediate mass host star in the same sense, was a remarkable (and lucky) break-through. Interpretation of the contemporaneous discovery of a faint point source around the famous debris disk host star Fomalhaut[141] has turned out to be more complex than anticipated[142]. Finally, at the end of 2008, a giant planet was found surrounding β Pictoris[120,143-145] within the prototypical debris disc first imaged in the early eighties[146]. These discoveries were preceded by several others, some of which were spurious, often around very young objects still in the process of becoming a star. For example, the companion to 2MASS 1207 (a very young brown dwarf) was discovered[147] through adaptive optics assisted near-infrared imaging on the *NACO* instrument on the *VLT*. This discovery was notable in three aspects: i) the system is very young, making detection of a self-luminous planetary mass object easier; ii) the central host object is of very low mass and thus of modest luminosity relative to the planetary mass companion; and iii) *NACO* is equipped with an infrared wave front sensor important to enable observations of this class of cool primaries. However, the mass ratio ($q$) of the brown dwarf to the companion is consistent with many examples of binary star systems of higher mass. To date, there are 10 objects found within 100 au of their host with a mass ratio between the companion and its host star $q < 0.02$ (Figure 3, http://exoplanet.eu/). These restrictions suggest they may have formed like planets in our Solar System, but this is not at all certain. There are dozens of objects that either have larger mass ratios (particularly around very low mass primaries) as well as objects with low mass ratios but found at larger radii (out to more than 1000 au). One major caveat to these studies is that the masses are inferred from theoretical models[148] based on the shape of the spectral energy distribution and (distance-dependent) luminosity, as well as knowledge about the central star (primarily age, but also composition).

The state-of-the-art instruments today require advanced adaptive optics to correct for the blurring effects of Earth's atmosphere[149]. While the diffraction limit improves at shorter wavelengths, high performance adaptive optics (AO) is more challenging, leading to compromises for instrument design between 0.5–5.0 µm. Even at the diffraction limit of an 8-meter-class telescope, one can only reach orbital separations of 3 au at 1.65 µm wavelength for stars out to 50 pc distance. The younger a planet is, the hotter and brighter it is, easing its detection and characterization. Nearby stars tend to be old (1 - 3 Gyr) and the youngest objects - being more rare - are located at greater distances. Thus another compromise needs to be found between available target sample and ease of detection, which translates directly into a balance between detectable mass (better for younger, more distance objects) and orbital separation (better for nearby stars). Results to date suggest that massive gas giant planets (> 2 $M_{♃}$) are rare at large orbital radii[150] (e.g. beyond 50 au). However, new instruments utilizing extreme adaptive optics (meaning an increase of hundreds to thousands of actuators controlling the shape of the deformable mirror, e.g., *SPHERE*[151] and *GPI*[152], will improve the inner working angle that can be reached at all wavelengths of operation, though in particular opening up the possibility of Strehl ratios above 30% in the red visible[153]. It is also worth mentioning that great improvements in data acquisition modes and analysis software (differential imaging via angular, polarimetric and spectral difference) have greatly enhanced planet-detection capabilities[154-158]. In addition, development of diffraction suppression optics continues as observations are contrast-limited close to the star. In the photon-noise limit, not often reached even around early-type bright stars, sparse aperture masking[159] and coronography can improve the achievable contrast limit using techniques such as apodizing phase plates[160,], vector vortex[161], phase-induced amplitude apodization[162], and classical Lyot coronography[163]. Dramatic improvements in diffraction suppression, stability and quality of adaptive optics, as well as in post-processing algorithms, are needed in order to reach the fundamental background-limited sensitivity close to the diffraction limit. The background limit is typically reached at 10 times larger inner working angles. The implementation of low-noise infrared wave-front sensors is another key area of development, particularly in its application to imaging surveys of fainter lower-mass stars and brown dwarfs. Building the observational data to constrain the frequency of planets as a function of planet mass, orbital separation and primary-star mass, will provide powerful tests to theories of planet formation.

The *James Webb Space Telescope (JWST)* will launch in 2018 and provide powerful capabilities for direct imaging, including coronagraphy. All of its instruments will make great contributions to finding and characterizing extra-solar planets resolved from their host stars, including some of those already known today. In particular, its short-wavelength imager *NIRCam* will be able to detect planets down to the masses of Neptune and Uranus beyond 30 au around close-by stars. *NIRISS*, a complementary imager, will utilize a sparse aperture mask to detect bright companions below the diffraction limit at 1–2.3 μm wavelength. It will be particularly useful for surveys of very young stars where planetary companions will be brightest relative to the central star. *MIRI*, the long-wavelength camera/spectrograph on *JWST*, will provide additional characterization of planetary atmospheres from 5–28 μm, and *NIRSPEC* (1–5 μm) will be equipped with an integral field spectrograph capable of providing high-quality spectra of close companions.

While JWST will be the most powerful telescope ever in terms of infrared sensitivity, it will not provide enhanced spatial resolution compared to the current generation of 6-10 meter telescopes and will not provide unique capabilities for high-contrast imaging at inner working angles below 0.1”. Because we know the distribution of giant gaseous planets rises with orbital radius out to 3 au, and since massive gas giants are rare beyond 30 au, it is likely that most Jupiter-mass planets will be found at intermediate separations. The next generation of extremely large telescopes (ELTs) will enable us to cross the 10 au threshold in angular resolution of accessible targets, pushing the detectable separation down to 3 au and enabling vast synergies between Doppler velocimetry and astrometry. The *Large Binocular Telescope Interferometer* (*LBTI*) is the first optical telescope with effective resolution of a 22-meter baseline[164], though it is not a filled aperture thus limiting its sensitivity. The *European ELT* (*E-ELT*), with its aperture of 39 meter, will integrate a suite of imaging and spectroscopic instruments (*HARMONI*, *MICADO*, *METIS*, and eventually *EPICS*) to enable efficient imaging of extra-solar planets at diffraction-limited inner working angles below 0.1”. Similar instruments are planned for the two other ELT projects, i.e. the Thirty-Meter Telescope (*TMT*) and the Giant Magellanic Telescope (*GMT*). Considerably thought is devoted today to figure out how to reach the fundamental background limit when approaching the diffraction limit, which, along with sensitivity and spatial resolution of these ELTs, would represent a major breakthrough. The removal of so-called quasi-static speckles is the key, which can be achieved (in principle) through sophisticated calibration schemes for AO systems to enable the commanded removal of speckles, or equally sophisticated analysis of the wave-front sensor camera data and telemetry to analyse residual errors in post-processing[165]. In predicting the performance of these future telescopes there is a tension between a conservative approach, on one side, and the wish to achieve the ultimate limits, on the other. Either way, the ELTs will represent a huge break-through in the capacity to directly image planets around nearby stars. If the technological challenges are mastered, the *E-ELT* will have a reasonable chance of obtaining a direct image of a super-Earth within 1 au from the nearest stars[166].

In late 2013, the European Space Agency launched *GAIA*, which has the capability of reaching micro-arcsecond astrometric precision. This mission will open up a wide range of discovery space to detect motions in the plane of the sky due to the orbit of the host star and planet around a common centre of mass. As the precision will fall for fainter stars, *GAIA* will be sensitive to the lowest-mass planets around stars in the solar neighbourhood, but will detect hundreds - if not thousands - of gas giant planets within hundreds of parsecs[167]. This will further open up a synergistic possibility to directly image some of these objects, providing ground-truth on the models for their evolution. Ground-based astrometry will also play a role, as could *JWST* and other facilities. For instance, ground-based direct imaging is able to deliver astrometric measurements at 100 micro-arcsecond precision[168].

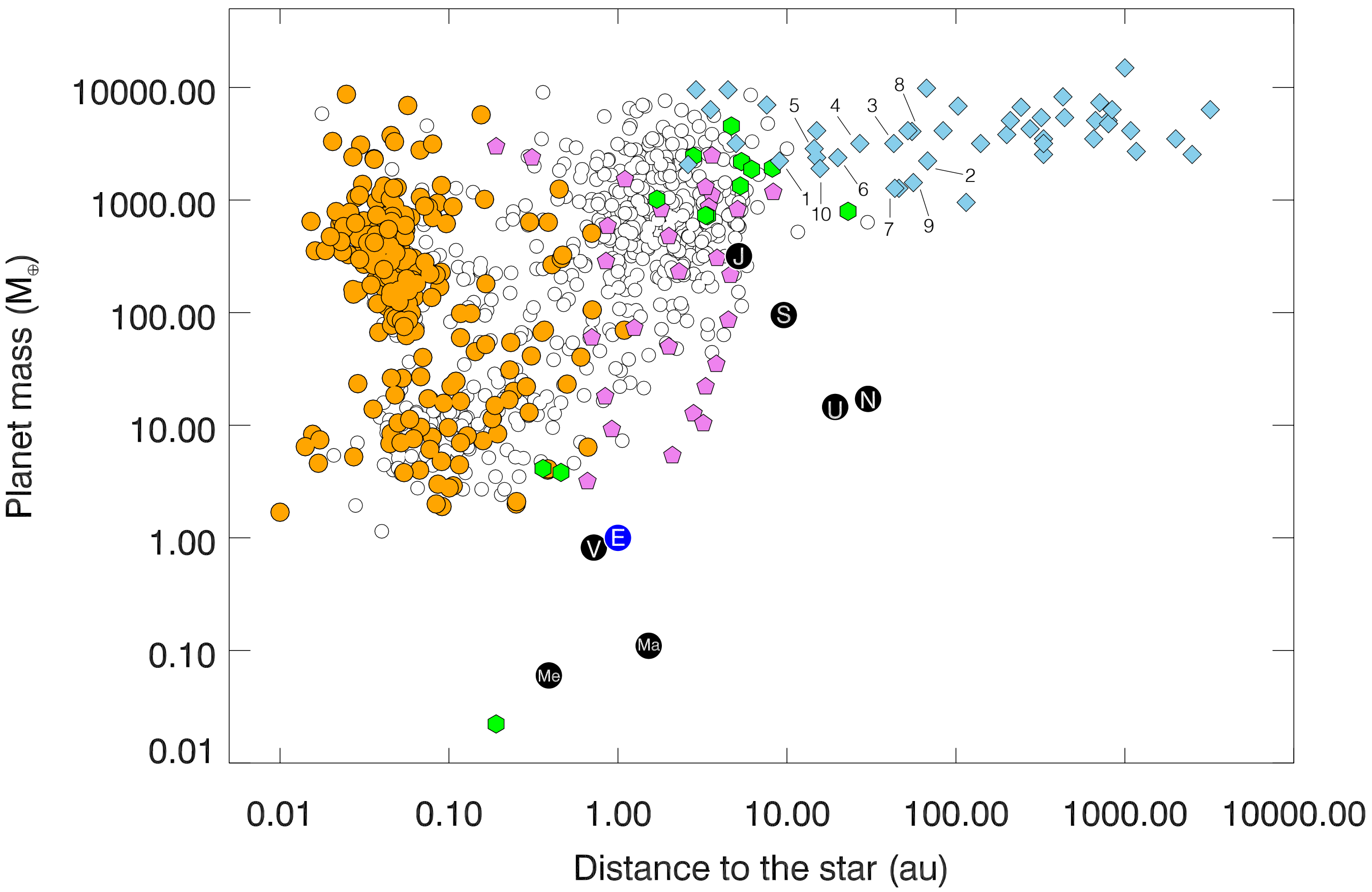

**Figure 3:** ***Mass and semi-major axis of known planets.*** **Planetary mass is plotted as a function of semi-major axis (the distance to the host star). Solar-system planets are shown by black circles, the Earth in blue. Exoplanets detected with different techniques and instrumentation are represented by different symbols: Doppler velocimetry (white circles), transit with a measured mass (orange circles), direct imaging (sky blue diamonds), microlensing (violet pentagons), and pulsation timing (green hexagons). Among the direct-imaging planets only ten were found within 100 au from their host and a mass ratio between the companion and its host star $q < 0.02$: beta Pic b, HR 8799e, PZ Tel b, HR 8799 d, HR 8799 c, GJ 504 b, kappa And b, HD 95086 b, HR 8799 b and LkCa 15b. Data underlying this plot were retrieved from the Exoplanet Encyclopaedia[196].**

## A bright and multi-technique future

In the past, focus was put on discovering new exoplanets and acquiring statistics about their diversity, which, in turn, concerned mainly external (orbital) parameters (see Figure 3 for a view on the current status of detection). Today, the interest is moving towards the detailed characterisation of specific planets and planetary systems. Orbital parameters, host star characteristics, synchronisation and planetary spin, irradiation, planet density and internal structure, atmospheric composition and physical conditions must be characterised in order to understand the formation processes and the observed diversity.

Increasing the number of targets amenable to further characterisation is the prime goal of several dedicated space mission projects: The extension of the *Kepler* mission (*K2*[169]), NASA's *Transiting Exoplanet Survey Satellite* (*TESS*[170]) and ESA's *Planetary Transits and Oscillations of Stars* mission

*(PLATO*[171]*)* will obtain photometric measurements of bright stars located almost everywhere on the sky, and thus find many new transiting planets around bright stars. These space missions will be complemented by new ground-based surveys dedicated to search for transits across different types of stars, e.g. the *Next Generation Transit Survey* (*NGTS*), the *Search for habitable Planets Eclipsing Ultra-Cool Stars* (*SPECULOOS*) and the *Exoplanets in Transit and Their Atmosphere* (*ExTrA*).

The planets transiting bright stars will enable follow-up observations and characterization of the planets by other techniques. The Swiss-ESA spacecraft *CHEOPS*[172,173] will, by transit photometry, measure precise radii and bulk densities of known exoplanets and select the best-suited targets for atmospheric characterisation with future spectrographs from space or on large ground-based telescopes. NASA's and ESA's *JWST* will enjoy of unprecedented thermal infrared sensitivity. Its four instruments will, in addition to direct imaging of planets, attempt transit observations at low- to medium-resolution ($100 < R < 1,500$) in the near- and mid-infrared domain for atmospheric characterisation. Whereas several of the known hot gas giants will be amenable to detail studies with *JWST* (Figure 2), additional low-mass targets, Neptunes, super-Earths and Earth-like planets, will be delivered by *TESS*, *CHEOPS*, and *PLATO*.

Atmospheric characterisation of transiting and non-transiting exoplanets has already been initiated with current ground-based direct imaging and resolved spectroscopy as well as high-resolution spectrographs (e.g. *CRIRES*, *HARPS*). These capabilities will be considerably extended with the upcoming generation of visible and near-infrared instruments equipping 4 to 8-m telescopes. The advent of extremely-large telescopes such as the *European ELT* (*E-ELT*), the *Thirty Meter Telescope* (*TMT*), and *Giant Magellan Telescope* (*GMT*), in combination with high spatial and spectral resolution, will amplify this tendency and open a new parameter space, for instance by enabling the detection of bands of molecular oxygen on super-earths transiting M dwarfs[174].

Radial-velocities, photometry, astrometry, imaging, spectroscopy, and other techniques will all contribute in a significant way to the field of exoplanets. Opposite to the past, they won't be in competition to each other but be highly complementary in view of a comprehensive understanding of the 'new worlds' mankind is eventually looking for. The more mature the field is becoming, the more it is understood that there won't be one single mission allowing us to find another Earth, but only the combination of all the tools being offered to us in the next decades.

**Table 1:** ***Non-exhaustive table of present (active) and future (approved) high-precision Doppler-velocimeters.*** **For the spectral band and the spectral resolution the maximum value is given. The (total) efficiency has been extrapolated to include slit losses, and telescope and atmospheric throughput. The radial-velocity precision was estimated from published orbits or standard star's velocities. Italic indicates design values and NA (Not Available) missing or non-reliable information. Historical instruments have not been listed. In is interesting to note that most of the planets discovered between 1995 and 2003, were detected using a small number of precision instruments: HIRES at the 10-m Keck i telescope in Hawaìi, CORALIE at the European Southern Observatory (ESO)'s 3.6-m telescope in La Silla, The Hamilton Spectrograph at the Shane 120-inch telescope at Lick, ELODIE at the 1.93-m telescope of the Haute-Provence Observatory[13], AFOE on the 1.5-m telescope at the Whipple Observatory[175], UCLES at the Anglo-Australian Telescope, Coudé Echelle Spectrograph[176] on the 2.7-m telescope, the Sandiford Cassegrain Echelle spectrograph[176] on the 2.1-m telescope both and HRS at the Hobby-Eberly Telescope, all of them at the McDonald Observatory. After 2003 the HARPS spectrograph opened a new window on the domain of super-Earths and mini-Neptunes by improving the radial-velocity precision below the meter-per-second level. Since, the meter-per-second precision has become a 'standard' and a goal for most Doppler-velocimeter projects presented in the Table.**

| Instrument/ Technique | Telescope/ Observatory | Start of operations | Band [µm] | Spectral resolution | Efficiency [%] | Precision [m s$^{-1}$] |
|---|---|---|---|---|---|---|
| HAMILTON[177] Self calibration | Shane 3-m Lick | 1986 | 0.34-1.1 | 30,000-60,000 | 3 - 6 | 3 |
| UCLES[178] Self calibration | 3.9-m AAT AAO | 1988 | 0.47-0.88 | - 100'000 | NA | 3 - 6 |
| HIRES[12] Self calibration | Keck I Mauna Kea | 1993 | 0.3-1.0 | 25,000-85,000 | 6 | 1 - 2 |
| CORALIE[13] sim. reference | EULER ESO La Silla | 1998 | 0.38-0.69 | 60,000 | 5 | 3 - 6 |
| UVES[179] Self calibration | UT2-VLT ESO Paranal | 1999 | 0.3-1.1 | 30,000 - 110,000 | 4 - 15 | 2 - 2.5 |
| HRS[180] Self calibration | HET McDonald | 2000 | 0.42-1.1 | 15,000 - 120,000 | 6 - 9 | 3 - 6 |
| HDS[181] Self calibration | Subaru Mauna Kea | 2001 | 0.3-1.0 | 90,000 - 160,000 | 6 - 13 | 5 - 6 |
| HARPS[18] Sim. reference | 3.6-m ESO La Silla | 2003 | 0.38-0.69 | 115,000 | 6 | < 0.8 |
| FEROS-II[182] Sim. reference | 2.2-m ESO La Silla | 2003 | 0.36-0.92 | 48,000 | 20 | 10 - 15 |
| MIKE[183] Self calibration | Magellan II Las Campanas | 2003 | 0.32-1.00 | 65,000 - 83,000 22,000 – 28,000 | 20 - 40 | 5 |
| SOPHIE[184] Sim. reference | 1.93-m OHP | 2006 | 0.38-0.69 | 39,000/75,000 | 4/8 | 2 |
| CRIRES[185] Self calibration | UT1-VLT ESO Paranal | 2007 | 0.95-5.2 | - 100,000 | 15 | 5 |
| PFS[186] Self calibration | Magellan II Las Campanas | 2010 | 0.39 – 0.67 | 38,000 – 190,000 | 10 | 1 |
| PARAS[187] Sim. reference | 1.2-m Mt. Abu | 2010 | 0.37-0.86 | 63,000 | NA | 3 - 5 |
| CAFE[188] Sim. reference | 2.2-m Calar Alto | 2011 | 0.39 – 0.95 | ~ 67,000 | 25 | 20 |
| CHIRON[189] Self calibration | 1.5-m CTIO | 2011 | 0.41-87 | 80,000 | 15 | < 1 |
| HARPS-N[50] Sim. reference | TNG ORM | 2012 | 0.38-0.69 | 115,000 | 8 | < 1 |
| LEVY[190] Self calibration | APF Lick | 2013 | 0.37-0.97 | 114'000-150'000 | 10 - 15 | < 1 |
| EXPERT-III[191] | 2-m AST | 2013 | 0.39 – 0.9 | 100'000 | NA | NA |
| NA | Fairborn | | | | | |
| GIANO[67] | TNG | 2014 | 0.95-2.5 | 50,000 | 20 | NA |
| Self calibration | ORM | | | | | |
| SALT-HRS[192] | SALT | 2014 | 0.38-0.89 | 16,000 - 67,000 | 10 - 15 | 3 - 4 |
| Self calibration | SAAO | | | | | |
| FIRST[191] | 2-m AST | 2014 | 0.8 – 1.8 | 60,000 - 72,000 | NA | NA |
| NA | Fairborn | | | | | |
| IRD[69] | Subaru | 2014 | 0.98-1.75 | 70,000 | NA | 1 |
| Sim. reference | Mauna Kea | | | | | |
| NRES | 6 x 1-m | 2015 | 0.39-0.86 | 53,000 | NA | 3 |
| NA | LCOGT | | | | | |
| MINERVA | 4 x 1-m | 2015 | 0.39-0.86 | NA | NA | 1 |
| Self calibration | Mt. Hopkins | | | (Kiwispec) | | |
| CARMENES[68] | Zeiss 3.5-m | 2015 | 0.55-1.7 | 82,000 | 10 - 13 | 1 |
| Sim. reference | Calar Alto | | | | | |
| PEPSI[193] | LBT | NA | 0.38-0.91 | 120,000 -320,000 | 10 | NA |
| Sim. reference | Mt. Graham | | | | | |
| HPF[70] | HET | NA | 0.98-1.40 | 50,000 | 4 | 1 - 3 |
| Sim. reference | McDonald | | | | | |
| CRIRES+ | VLT | 2017 | 0.95-5.2 | - 100,000 | 15 | < 5 |
| Self calibration | ESO Paranal | | | | | |
| ESPRESSO[194] | All UTs- | 2017 | 0.38-0.78 | 60’000 - 200,000 | 6 - 11 | 0.1 |
| Sim. reference | VLT | | | | | |
| | ESO Paranal | | | | | |
| SPIROU[72] | CFHT | 2017 | 0.98-2.35 | 70,000 | 10 | 1 |
| Sim. reference | Mauna Kea | | | | | |
| G-CLEF[195] | GMT | 2019 | 0.35-0.95 | 120,000 | 20 | 0.1 |
| Sim. reference | Las Campanas | | | | | |

## ACKNOWLEDGEMENTS

DEH would like to dedicate this article to the memory of STIS Principal Investigator, Bruce Woodgate, who passed away on April 2014. This work has been carried out within the frame of the National Centre for Competence in Research "PlanetS" supported by the Swiss National Science Foundation (SNSF). The authors acknowledge the financial support of the SNSF.

## AUTHOR CONTRIBUTION

All the co-authors participated to the writing effort of the review article. More specifically, FPE contributed to section 2 on radial velocities, DEH to section 3 and 4 about transit observations, microlensing and the characterization of planetary atmospheres, MME to section 5 about gas giant formation, imaging and how to probe the outer disk.

## AUTHOR INFORMATION

Reprints and permissions information is available at ???. Correspondence and requests for materials should be addressed to Francesco.Pepe@unige.ch.

Francesco Pepe, FPE, Francesco.Pepe@unige.ch
David Ehrenreich, DEH, David.Ehrenreich@unige.ch
Michael R. Meyer, MME, mmeyer@phys.ethz.ch